\documentclass[reprint, amsmath,amssymb, pre]{revtex4-1}
\usepackage{graphicx}
\usepackage{dcolumn}
\usepackage{bm}
\usepackage{amsmath}
\usepackage{mathrsfs}
 \usepackage{amssymb}
\usepackage{algorithm}
\usepackage{algorithmic}
\usepackage{color}
\begin{document}

\title{Spin-glass model for the C-dismantling problem}

\author{Shao-Meng Qin}
\email{qsminside@gmail.com}
\affiliation{College of Science, Civil Aviation University of China, Tianjin 300300,  China}
\date{\today}

\begin{abstract}
$C$-dismantling (CD) problem aims at finding the minimum vertex set $\mathcal{D}$ of a graph $\mathcal{G}(\mathcal{V},\mathcal{E})$ after removing which the remaining graph will break into connected components with the size not larger than $C$.
In this paper, we introduce a spin-glass model with $C+1$ integer-value states into the CD problem and then study the properties of this spin-glass model by the belief-propagation (BP) equations under the replica-symmetry ansatz.
We give the lower bound $\rho_c$ of the relative size of $\mathcal{D}$ with finite $C$ on regular random graphs and  Erd\"os-R\'enyi random graphs.
We find $\rho_c$ will decrease gradually with growing $C$ and it converges to $\rho_\infty$ as $C\to\infty$.
The CD problem is called dismantling problem when $C$ is a small finite fraction of $|\mathcal{V}|$.
Therefore, $\rho_\infty$ is also the lower bound of the dismantling problem when $|\mathcal{V}|\to\infty$.
To reduce the computation complexity of the BP equations, taking the knowledge of the probability of a random selected vertex belonging to a remaining connected component with the size $A$, the original BP equations can be simplified to one with only three states when $C\to \infty$.
The simplified BP equations are very similar to the BP equations of the feedback vertex set spin-glass model [H.-J.~Zhou, Eur.~Phys.~J.~B {\bf 86}, 455 (2013)].
At last, we develop two practical belief-propagation-guide decimation algorithms based on the original BP equations (CD-BPD) and the simplified BP equations (SCD-BPD) to solve the CD problem on a certain graph.
Our BPD algorithms and two other state-of-art heuristic algorithms are applied on various random graphs and some real world networks.
Computation results show that the CD-BPD is the best in all tested algorithms in the case of small $C$. But considering the performance and computation consumption, we recommend using SCD-BPD for the network with small clustering coefficient when $C$ is large.
\end{abstract}

\pacs{Valid PACS appear here}
\maketitle
\section{Introduction}
In a graph $\mathcal{G}(\mathcal{V},\mathcal{E})$ with $N=|\mathcal{V}|$ vertices and $M=|\mathcal{E}|$ undirected edges, there exist some vertices which are crucial to the connectivity of the graph.
The set of these vertices $\mathcal{D}$ is called $C$-dismantling (CD) set if its removal yields a remaining graph in which the size of each connected component will be equal to or smaller than $C$~\cite{EDWARDS200130,janson_thomason_2008}.
In the past few years, researchers work on the topic of the CD problem which asks finding the minimum CD set  of a graph, especially in the case of $C$ taking a finite fraction of $N$, like $C/N=0.01$, which is also named as dismantling problem \cite{Braunstein2016,Mugisha2016,Zhou2013,Zdeborova2016,Morone2015,Clusella2016,Ren2018Generalized, PhysRevE.98.012313}.

For some real world networks, such as transportation network and internet, their robustness and function depend on their scale and connectivity to a large extent \cite{PhysRevLett.85.5468,Cohen2001,Albert2000}.
On the other side, we can also stop epidemic (or computer virus) spreading by vaccinating the people (or computer) who can divide the infection network to separated components \cite{Guggiola2015,PhysRevLett.86.3200,PhysRevX.4.021024}.
Therefore, as one of the fundamental problems in the network science, the CD problem relates to many other important problems and practical applications, ranging from the percolation problem \cite{Zhang2017}, to the information spreading \cite{WeiChen, Sun2016} and so on.

In 2008, Janson and Thomason proved some useful properties of the CD problem on sparse random graphs~\cite{janson_thomason_2008}.
But the problem of finding the minimum CD set or dismantling set of a certain graph belongs to the nondeterministic polynomial hard (NP-hard) class of computational complexity \cite{Zdeborova2016,Karp1972,Lu2016}.
Therefore, researchers are not pinning their hopes on solving this problem by a complete algorithm in time bounded by a polynomial function of $M$ or $N$ but devote their efforts to all kinds of heuristic methods.
The starting points of these heuristic algorithms, even without strict proof but whose rationality has been proved by their pretty results in solving dismantling problem, are the possible correlations between network structure and network attacking.
Except some methods based on the vertices' highest degree \cite{Albert2000,PhysRevLett.85.5468,Xia2017} or betweenness \cite{PhysRevE.65.056109}, Morone and Makse considered the information spreading of vertices and proposed the collective information algorithm \cite{Morone2015}.
The authors claimed that it beats all existing heuristic algorithms at that time.
Another algorithm recursively removes vertices having the highest degree from the 2-core of the graph, which is obtained by adaptive removal of all leaves \cite{Zdeborova2016}.
The CD problem can also be solved by the node explosive percolation algorithms, which start from a completely dismantled graph and then reconstructs the graph by adding the removed vertex back \cite{Schneider2012,Clusella2016,PhysRevE.98.012313}.
Later, some other researchers pointed out that for random graphs the dismantling problem is equivalent to the decycling problem, which is also referred as the feedback vertex set (FVS) problem and aims to remove as few vertices as possible to break all cycles in the graph~\cite{Zhou2016,Zhou2013,Qin2014,PhysRevE.94.022146,Guggiola2015}.
The dismantling algorithms stemmed from the FVS problem are characterized by their perfect performance in giving a very small dismantling set and their results are very close to the theoretically optimal value of the decycling problem \cite{Zdeborova2016,Braunstein2016,Mugisha2016}.

Most of studies discussed above mainly focus their attention on the dismantling problem, where $C$ will be very large as $N\to\infty$, but they cannot guarantee their performance in the case of small $C$.
In order to have a more comprehensive understanding of the CD problem and to solve this problem efficiently, we focus our attention on the general CD problem with finite $C$ and propose a spin-glass model  with $C+1$ states to describe the constraints in the CD problem.
Considering that the removal cost of a vertex depends on the protection effort on it, we introduce the spin-glass model considering the situation where each vertex has variable removal costs.
By using the belief-propagation (BP) equations under the replica-symmetry (RS) ansatz, we study various properties of the CD problem, including the lower bound of the relative size of minimum $\mathcal{D}$ with $C$, the probability of a random selected vertex belonging to a connected component with size $A$, the complexity of the BP equations and the connection between CD problem and the FVS problem.
What is more, we also develop two belief-propagation-guide decimation algorithms (CD-BPD and SCD-BPD) to solve the CD problem on a certain graph.
The CD-BPD algorithm is based on BP equations of our spin-glass model and the SCD-BPD is a coarse-gained algorithm of the CD-BPD in large $C$ limit.
Our extensive numerical computations on artificial random graphs and real world networks exhibit that CD-BPD has significant advantages in solving CD problem over others in the case of small $C$.
When $C$ is large, the CD-BPD or SCD-BPD is also the best algorithm for the CD problem on networks with small clustering coefficient~\cite{Watts1998}.

This paper is organized as follows. In the next section we introduce the spin-glass model for the CD problem and explain how to compute thermodynamic quantities under the RS ansatz.
The numerical computation results on random graphs and some real world networks are given in the Sec.~\ref{sec:result}.
In the last section, we conclude our work and discuss some possible extensions.

\section{Spin-Glass Model of the CD problem}

After a graph has been $C$-dismantled, the remaining graph will break into numerous connected components not larger than $C$. Therefore the minimum CD problem is completely equivalent to the maximum $C$-component set problem asking the maximum set of vertex $\mathcal{S}=\mathcal{V}\setminus \mathcal{D}$, so that the vertices in $\mathcal{S}$ and the edges between them form connected components not larger than $C$.
In the case of $C=1$, $C$-component set problem is equivalent to another NP-hard problem: the vertex cover problem, which has been analyzed by the RS mean-field method extensively \cite{PhysRevE.97.022138, Weigt2006,PhysRevE.94.022146,ZJH78901,PhysRevE.63.056127}.
Inspired by the spin-glass model of the vertex cover problem, in this paper, we develop our spin-glass model for the $C$-component set problem as well as the CD problem and then analyze its properties by the RS mean-field method.

In a dismantled graph, if vertex $i\in\mathcal{S}$ is in the connected component $\mathcal{C}_\alpha$, we use an integer-value $A_i=|\mathcal{C}_\alpha|$ to present the state of vertex $i$ in our spin-glass model, where $|\mathcal{C}_\alpha|$ means the size of $\mathcal{C}_\alpha$ or the number of vertices in $\mathcal{C}_\alpha$.
If the vertex $i\in\mathcal{D}$, we say the vertex $i$ is in a connected component with size 0 and $A_i=0$.
In CD problem, the size of each remaining connected component must be equal to or smaller than $C$, so $A_i$ can only take $C+1$ different integer-values from 0 to $C$.
A microscopic configuration $\underline{A}\equiv\{A_1,A_2,\cdots,A_N\}$ of graph $\mathcal{G}$ is called legitimate if and only if the following constraint is fulfilled
\begin{equation}
L(\underline{A})\equiv\prod_i \delta(A_i,\Gamma(i)),
\end{equation}
where $\Gamma(i)$ returns the size of the connected component containing vertex $i$ in the dismantled graph, and $\delta(x,y)$ is the Kronecker delta function such that $\delta(x,y)=1$ if $x=y$, and $\delta(x,y)=0$ if $x\neq y$.

In this spin-glass model, if we consider the removal cost $\omega_i\geq 0$ of each vertex $i$,
the CD problem should pursuit for the minimum total removal cost instead of the minimum $\mathcal{D}$.
Therefore, the energy of $\underline{A}$ is defined as the total removal cost of all vertices with $A_i=0$,
\begin{equation}
\label{eq:E}
E(\underline{A})=\sum_i \omega_i \delta(A_i,0).
\end{equation}

We assume the spin-glass system follows the Boltzmann distribution and the probability of observing a legitimate state $\underline{A}$ is
\begin{equation}
\label{eq:pA}
p(\underline{A})=\frac{\exp(-\beta E(\underline{A}))}{Z(\beta)},
\end{equation}
where $\beta$ is the inverse temperature in the canonical ensemble and $Z(\beta)$ is the partition function
\begin{equation}
\label{eq:Z}
Z(\beta)=\sum_{\underline{A}} \exp(-\beta E(\underline{A}))L(\underline{A}).
\end{equation}

Now we consider the marginal probability of a vertex $i$ taking the state $A_i$, denoted as $q_i^{A_i}$.
The value of $q_i^{A_i}$ is strongly influenced by the marginal probabilities of $i$'s nearest neighbours $j\in \partial i$, where $\partial i$ gives the set of nearest neighbour vertices of $i$ in graph $\mathcal{G}$.
After we build a cavity graph $\mathcal{G}_{\setminus i}$ by removing vertex $i$ from $\mathcal{G}$,
we can use the Bethe-Peierls approximation to neglect all possible correlations among the marginal probabilities of vertices $j\in\partial i$ \cite{Peierls1936,Bethe1935}, which is denoted as $q_{j\to i}^{A_j}$.
Then we can have the value of $q_i^{A_i}$ by the following equations:
\begin{subequations}
\label{eq:RS}
  \begin{align}
q_i^0&=\frac{e^{- \beta\omega_i}}{z_i}\;,\label{eq:RS0}\\
q_i^{A_i}&=\frac{1}{z_i}\sum_{\underline{A}_{\partial i}}\delta(\sum\limits_{j \in\partial i} A_j+1,A_i)\prod_{j \in\partial i} q_{j\rightarrow i}^{A_j}\quad  (A_i\neq 0) \;,
\end{align}
\end{subequations}
where $\underline{A}_{\partial i}\equiv\{A_j\}_{j\in\partial i}$ is the local configuration of vertex $i$ and the normalization factor $z_i$ is
\begin{equation}
\label{eq:zi}
z_i\equiv e^{- \beta\omega_i}+\sum_{\underline{A}_{\partial i}}H(C-\sum\limits_{j \in\partial i} A_j-1)\prod_{j \in\partial i} q_{j\rightarrow i}^{A_j},
\end{equation}
where $H(x)$ is the Heaviside step function such that $H(x)=1$ if $x\geq 0$, and $H(x)=0$ if $x<0$.

Equation~\ref{eq:RS} considers constraints that only when the total size of all neighbour components is smaller than $C-1$, we can accept the vertex $i\in\mathcal{S}$ which will merge all neighbour components together to a bigger one with the size $\sum_{j\in \partial i}A_j+1$.
If $\sum_{j\in \partial i} A_j\geq C$, the vertex $i$ must be in set $\mathcal{D}$ to prevent forming a connected component whose size exceeds $C$. $q_{i\rightarrow j}^{A_i}$ has the same meaning with $q_i^{A_i}$ except that it is defined on the cavity graph $\mathcal{G}_{\setminus i}$. The self-consistency BP equations of $q_{i\rightarrow j}^{A_i}$ is
\begin{subequations}
\label{eq:BP}
  \begin{align}
q_{i\rightarrow j}^0&=\frac{e^{- \beta\omega_i}}{z_{i\to j}}\;,&\\
q_{i\rightarrow j}^{A_i}&=\frac{1}{z_{i\to j}}\sum_{\underline{A}_{\partial i\setminus j}}\delta(\sum\limits_{k \in\partial i\setminus j} A_k+1,A_i)&\prod_{k \in\partial i\setminus j} q_{k\rightarrow i}^{A_k}\nonumber\\
&&(A_i\neq 0)\;,
\end{align}
\end{subequations}
where $\partial i\setminus j$ means the vertex set obtained by deleting vertex $j$ from $\partial i$ and the normalization factor $z_{i\rightarrow j}$ is
\begin{equation}
z_{i\rightarrow j}\equiv e^{- \beta\omega_i}+\sum_{\underline{A}_{\partial i\setminus j}}\prod_{k \in\partial i\setminus j} q_{k\rightarrow i}^{A_k}H(C-\sum_{k \in\partial i\setminus j} A_k-1).
\end{equation}

For a certain graph instance $\mathcal{G}$, the BP equations can be solved by iterating the equations on edges at a fixed $\beta$.
After BP equations are solved, we can obtain thermal dynamical quantities of this spin-glass system under the Bethe-Peierls approximation.
We start from the free energy $F=\sum_i f_i-\sum_{(i,j)\in G}f_{ij}$, where the $f_i$ and $f_{ij}$ are free energy contribution from vertex $i$ and edge $(i,j)$,
\begin{eqnarray}
\label{eq:f}
f_i&= &-\frac{1}{\beta}\ln z_i,\\
f_{ij}&=&-\frac{1}{\beta}\ln \sum_{A_i,A_j} H(C-A_i-A_j) q_{i\rightarrow j}^{A_i}q_{j\rightarrow i}^{A_j}.
\end{eqnarray}
As free energy is an extensive quantity, we are more interesting to free energy density obtained by $f=F/N$.
The energy density $e$ of the spin-glass model equals to
\begin{equation}
e=\frac{<E>}{N}=\frac{1}{N}\sum_i  q_i^0 \omega_i.
\end{equation}
If all vertices have uniform removal cost and $\omega_i=1$, the energy density is the relative size of the set $\mathcal{D}$.
At last, we can obtain the entropy density by
\begin{equation}
\label{eq:entropy}
s=\beta(e-f).
\end{equation}
\section{Results}
\label{sec:result}
We now apply the CD spin-glass model on the regular random graphs (RR),  Erd\"os-R\'enyi (ER) random graphs, scale-free (SF) graphs and some real world networks.
In the present paper, we generate the SF networks by a static method explained in \cite{PhysRevLett.87.278701}.
Because we do not have the knowledge of the removal cost,  we assume the removal cost $\omega_i=1$ for all vertices.
Actually, the BP equations still hold even if the removal cost is not uniform.
Without being specific, the results of artificial random graphs in the following discussion are obtained by averaging over 16 different instances with $N=2^{17}$.

\subsection{Statistical properties of the CD problem on random graphs}
Equation \ref{eq:Z} tells us that the partition function will be dominated by low-energy configurations when $\beta$ is large.
What is more, in most spin-glass systems including this one, the number of configurations will decrease with decreasing energy and the entropy density $s$ will also decrease with growing $\beta$.
As the entropy of a real system must be nonnegative, the mean-field result predicts the relative size of the minimum $\mathcal{D}$, denoted by $\rho_c$, at the inverse temperature $\beta=\beta^*$ where $s(\beta^*)=0$.
For RR graph, each vertex has the same degree, so we can have the numerical solution of the BP equations~\ref{eq:BP}.
For the ER graph, we use population dynamics to investigate $\rho_c$ with various average degree and $C$ \cite{LenkaThe,MezardM2009, Mezard2001,PhysRevE.77.066102}.

\begin{figure*}
  \includegraphics[width=0.80\textwidth]{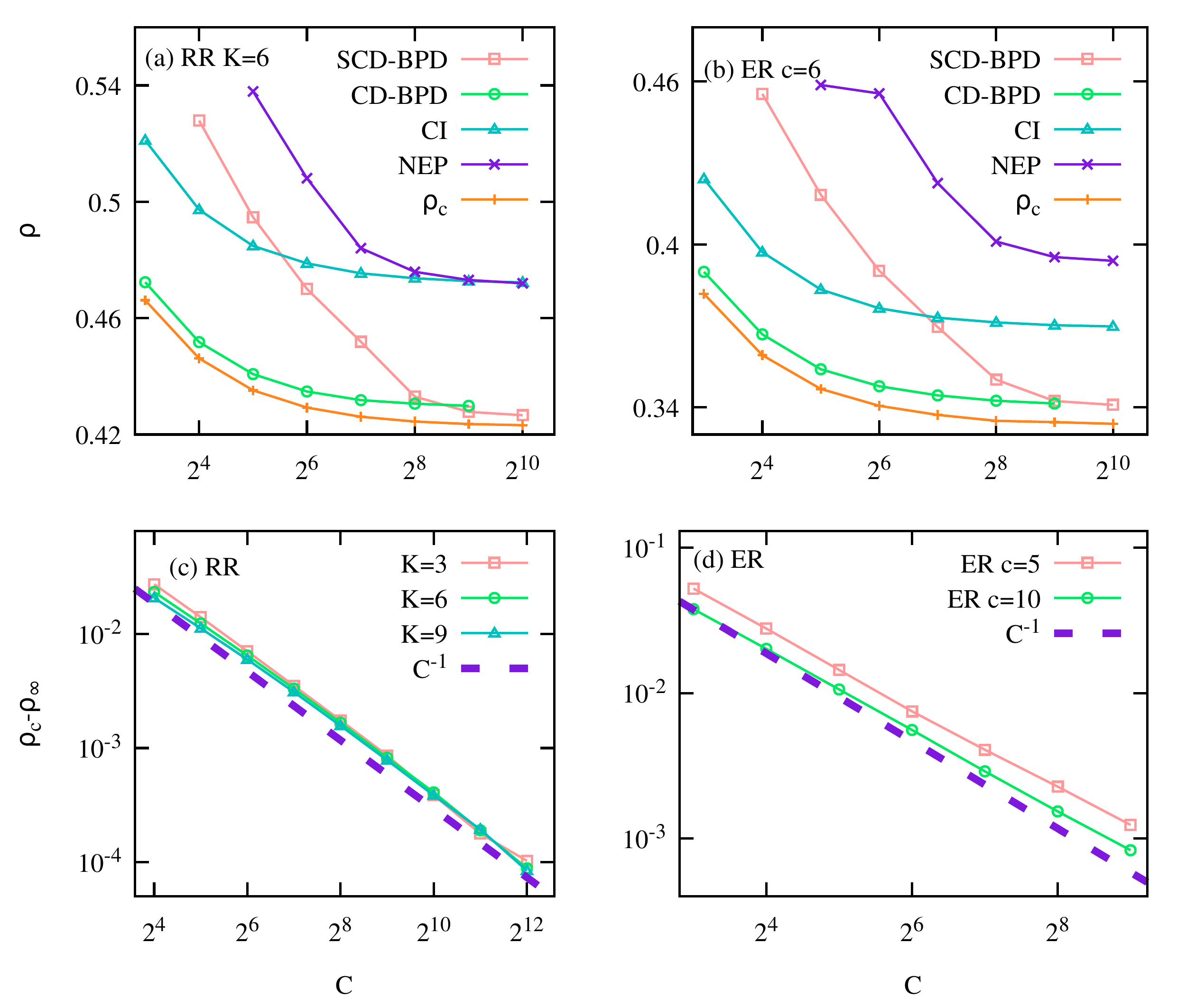}
  \caption{(Color online) (a) Fraction of vertices removed as a function of $C$ on RR graph with degree $K=6$ given by the RS mean-field method (pluses), the CD-BPD (circles), the SCD-BPD (squares) algorithms, the CI (triangles) and the NEP (crosses). (b) The same as (a) but on ER graph with average degree $c=6$. (c) Scaling plot of the $(\rho_c-\rho_\infty)$ versus $C$ for RR graph with $K=3$ (squares) where $\rho_\infty=0.24236$, $K=6$ (circles) where $\rho_\infty=0.42278$ and $K=9$ (triangles) where $\rho_\infty=0.519633$. (d) The same as (c) but on ER graph with degree $c=5$ (squares) where $\rho_\infty=0.2785$ and $c=10$ (circles) where $\rho_\infty=0.4835$.
    \label{fig:fig1}
  }
\end{figure*}

\begin{figure*}
  \includegraphics[width=0.98\textwidth]{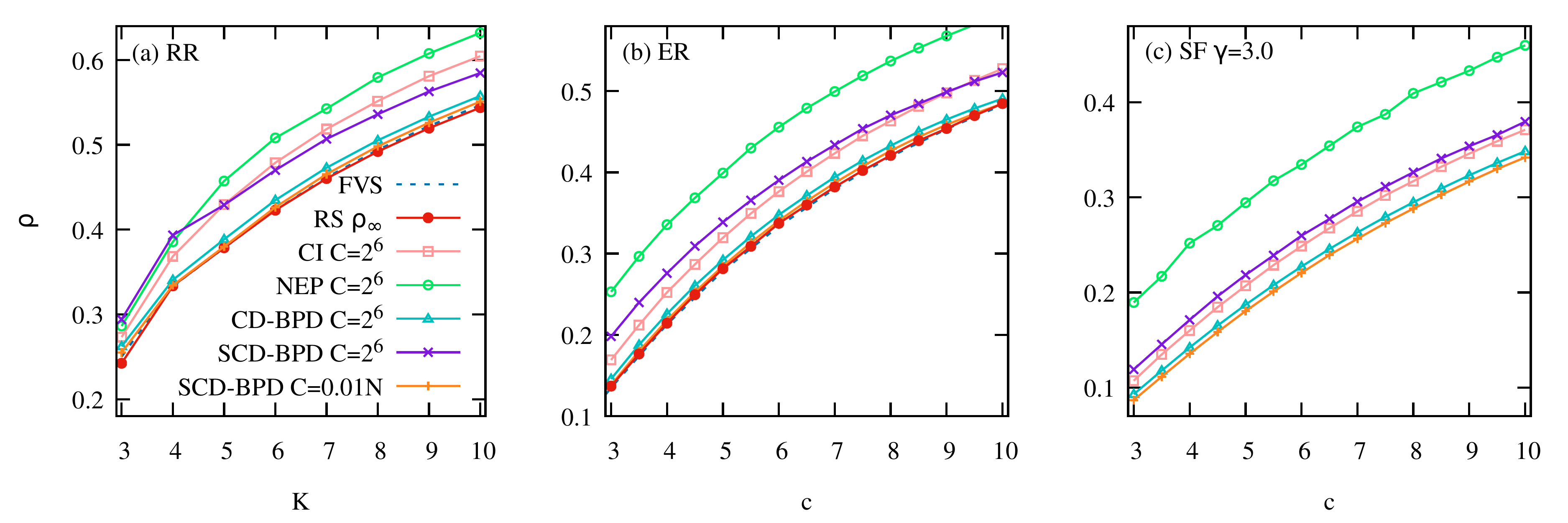}
  \caption{ (Color online) The relative size $\rho$ of set $\mathcal{D}$ for (a) RR graph on degree K, (b) ER graph on mean degree c and (c) SF graph on mean degree c with power-law exponent $\gamma=3.0$ given by collective influence algorithm with ball radius $\ell =2$ (CI) (squares)~\cite{Morone2015}, node explosive percolation algorithm with the second score definition in~\cite{Clusella2016} (NEP) (empty circles), CD-BPD (triangles) and SCD-BPD (crosses) algorithms.
  $\rho_\infty$ is the value of $\rho_c$ at $C\to\infty$ predicted by the RS mean-field method (solid circles).
  The dashed lines are the lower bounds of the minimum FVS predicted by the RS mean-field method \cite{Zhou2013}.
  \label{fig:fig4}
  }
\end{figure*}

From Fig.~\ref{fig:fig1}(a) and (b), we can see $\rho_c$ decreases gradually with $C$ until it converges to $\rho_\infty$ as $C\to\infty$.
Finite-size scaling analysis in Fig.~\ref{fig:fig1}(c) and (d) exhibits $(\rho_c-\rho_\infty)\propto C^{-\zeta}$.
The exponent $\zeta\approx 1$ is almost irrelevant with the degree distribution of random networks. Because of the locally tree-like character of random graphs,  $\zeta\approx 1$ agrees with the exponent of tree network \cite{janson_thomason_2008}.
The value of $\rho_\infty$ can also be obtained by extrapolating the result of $\rho_c$ in the large $C$ limit.
As discussed above, the dismantling problem can be regarded as the CD problem with infinite $C$.
Therefore, $\rho_\infty$ gives the fraction of the minimum removed vertices in dismantling problem.
The value of $\rho_\infty$ for RR and ER networks with various degree are presented in Fig.~\ref{fig:fig4}: $\rho_\infty$ increases monotonically with growing mean vertex degree in RR and ER random graph ensembles.
We also notice that the difference between the $\rho_\infty$ and the relative size of the minimum FVS predicted in \cite{Zhou2013} is inconspicuous, which will be explained in the following discussion.

\begin{figure}
  \includegraphics[width=0.35\textwidth]{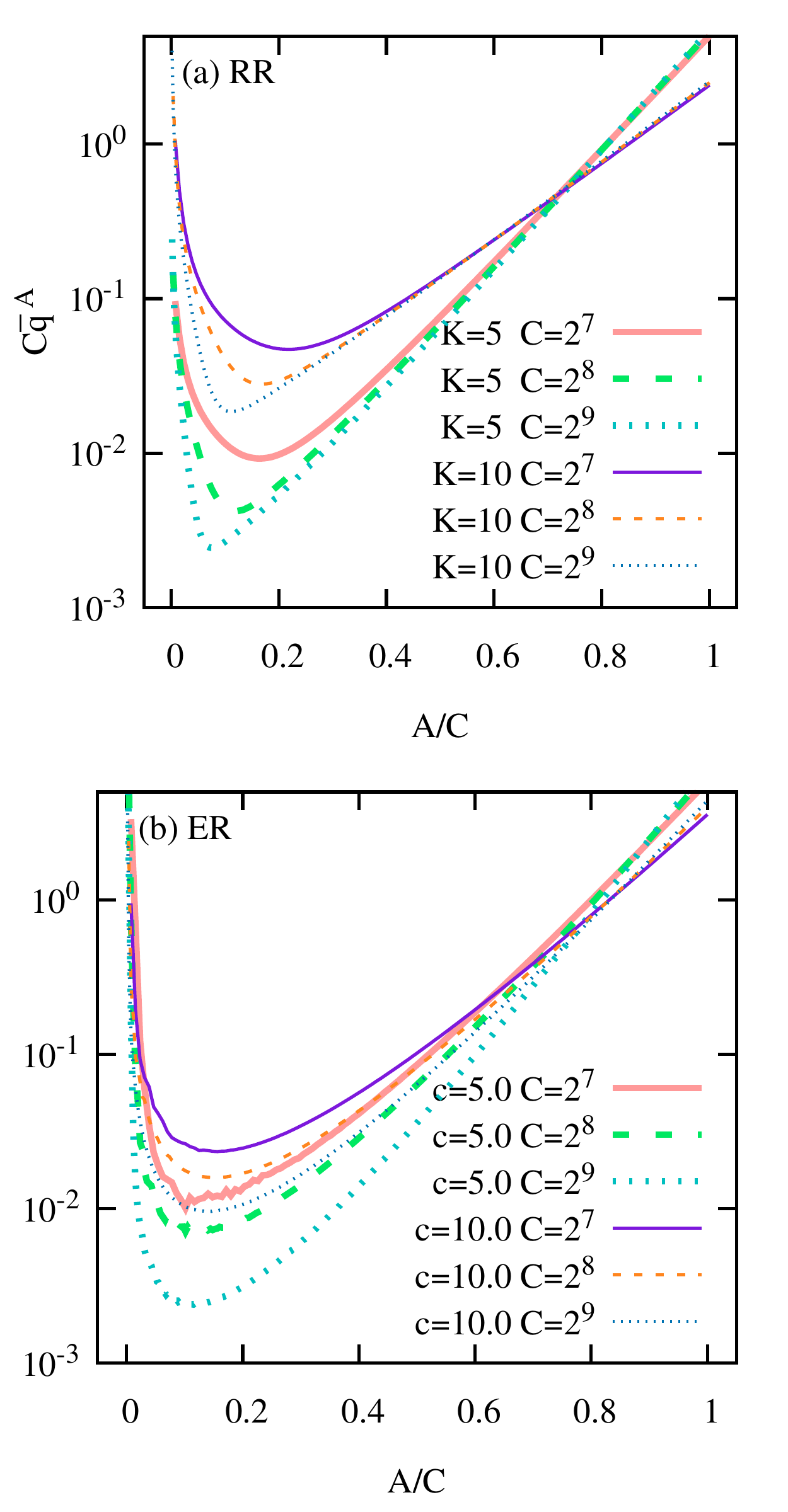}
  \caption{(Color online)
  The mean value of the probability $q_i^{A}$ over all vertices in (a) RR graph with degree $K=5$ and $C=2^7$ (thick solid line), $C=2^8$ (thick dashed line), $C=2^9$ (thick dotted line) and with degree $K=10$ and $C=2^7$ (thin solid line), $C=2^8$ (thin dashed line), $C=2^9$ (thin dotted line)  (b) ER graph with average degree $c=5$ and $C=2^7$ (thick solid line), $C=2^8$ (thick dashed line), $C=2^9$ (thick dotted line) and with degree $c=10$ and $C=2^7$ (thin solid line), $C=2^8$ (thin dashed line), $C=2^9$ (thin dotted line).
  In order to compare the probability distributions with different $C$ easily, these distributions are multiplied by $C$ respectively.
  $\overline{q}^{0}$ is not presented in these figures because it is far beyond the others.
    \label{fig:fig2}
  }
\end{figure}

The mean value of the $q_i^{A}$ over all vertices, denoted as $\overline{q}^A=\frac{1}{N}\sum_i q_i^{A}$, is the probability of a randomly selected vertex in a connected component with size $A$.
We compute the value of $\overline{q}^A$ for RR and ER graphs and present the results in Fig. \ref{fig:fig2}.
In the case of small $A$, $\overline{q}^A$ will decay with growing $A$ quickly.
But as $A$ approaching to $C$, all curves in Fig.~\ref{fig:fig2} increase with $A/C$ exponentially.
In the large $C$ limit, the exponents of these curves mainly depend on the type and the mean vertex degree of the graph.
Moreover, $\overline{q}^A$ reaches its minimum at the point $A_{\min}=\arg \min \overline{q}^A$ and $\lim_{C\to\infty}A_{\min}/C=0$.
Therefore, in the case of $C\to\infty$, there are two types of connected components in a dismantled graph: small connected components with a few vertices and large connected components with the size close to $C$.

\subsection{The computation complexity of the BP equations}

In this subsection, we will discuss the computation complexity of the Eq.~\ref{eq:RS} and \ref{eq:BP}.
The readers may argue that we must consider all local microscopic configurations $\underline{A}_{\partial i}$ to compute the message $q_{i\to j}^{A_i}$ in Eq. \ref{eq:RS}, so the computation complexity of the BP iterations must be larger than $(C+1)^{|\partial i|}$.
However, only the configurations with $\sum_{j \in\partial i} A_j+1<C$ work in the equations, which take a very small fraction of all local configurations.
The real computation complexity of the BP equations will also be much smaller than $(C+1)^{|\partial i|}$.

We start from neglecting all nearest neighbours of vertex $i$ except $m$ and $n$. If $m$ and $n$ are in connected components with the size $A_m$ and $A_n$ respectively in cavity graph $\mathcal{G}_{\setminus i}$ and $i\in\mathcal{S}$, two connected components will combine together to a new one with the size $A_m+A_n+1$.
The probability of $A_m+A_n$ can be described by a new introduced probability distribution $\tilde{q}^A$ in which $A\in\{0,\cdots,C\}$ and
\begin{subequations}
\label{eq:otimes}
  \begin{align}
\tilde{q}^A&=\sum_{A_m+A_n<C} q_{m\to i}^{A_m}q_{n\to i}^{A_n}\delta(A_m+A_n,A) \quad (A<C)\;,\\
\tilde{q}^C&=\sum_{A_m+A_n\geq C} q_{m\to i}^{A_m}q_{n\to i}^{A_n}\;.
\end{align}
\end{subequations}
Because the size of connected component cannot be larger than $C$, we only concern the situations of $A_m+A_n<C$ and $A=A_m+A_n$ in this case.
For the $A=C$, we sum all probabilities of $A_m+A_n\geq C$ together.
In our later discussion,  Eq.~\ref{eq:otimes} is abbreviated as $\tilde{q}^A=q_{m\to i}^{A_m} \otimes q_{n\to i}^{A_n}$.

Now we consider all nearest neighbours of vertex $i$. In the same way, we can use a product of $\otimes$ to compute the probability of $\tilde{q}_i^{A_i}$, which means the probability distribution of $A_i=\sum_{j\in\partial i} A_j$ when $A_i<C$ and $A_i=C$ when $\sum_{j\in\partial i} A_j\geq C$:
\begin{equation}
\tilde{q}_i^{A_i}=\prod_{k\in\partial i}^\otimes q_{k\to i}^{A_k}.
\end{equation}
In the following, we can obtain the value of $q_i^{A_i}$ easily from $\tilde{q}_i^{A_i}$:
\begin{subequations}
\label{eq:newRS}
  \begin{align}
q_i^0&=e^{-\beta\omega_i}\slash z_{i} \;,\\
q_i^{A_i}&=\tilde{q}_i^{A_i-1} \slash z_{i}\;\quad (A_i\neq 0),
\end{align}
\end{subequations}
where
\begin{equation}
z_i\equiv e^{-\beta\omega_i}+\sum_{A=0}^{C-1}\tilde{q}_i^A.
\end{equation}

Similarity, we can also compute the cavity message $q_{i\to j}^{A_i}$ from $\tilde{q}_{i\to j}^{A_i}$:
\begin{subequations}
\label{eq:newBP}
  \begin{align}
q_{i\to j}^0&=e^{-\beta\omega_i}\slash z_{i\to j} \;,\\
q_{i\to j}^{A_i}&=\tilde{q}_{i\to j}^{A_i-1} \slash z_{i\to j}\;\qquad (A_{i\to j}\neq 0),
\end{align}
\end{subequations}
where
\begin{eqnarray}
\tilde{q}^{A_i}_{i\to j}&=\prod\limits_{k\in\partial i\setminus j}^\otimes q^{A_k}_{k\to i},\\
z_{i\to j}&\equiv e^{-\beta\omega_i}+\sum\limits_{A=0}^{C-1}\tilde{q}_{i\to j}^A.
\end{eqnarray}

\begin{figure}
  \includegraphics[width=0.4\textwidth]{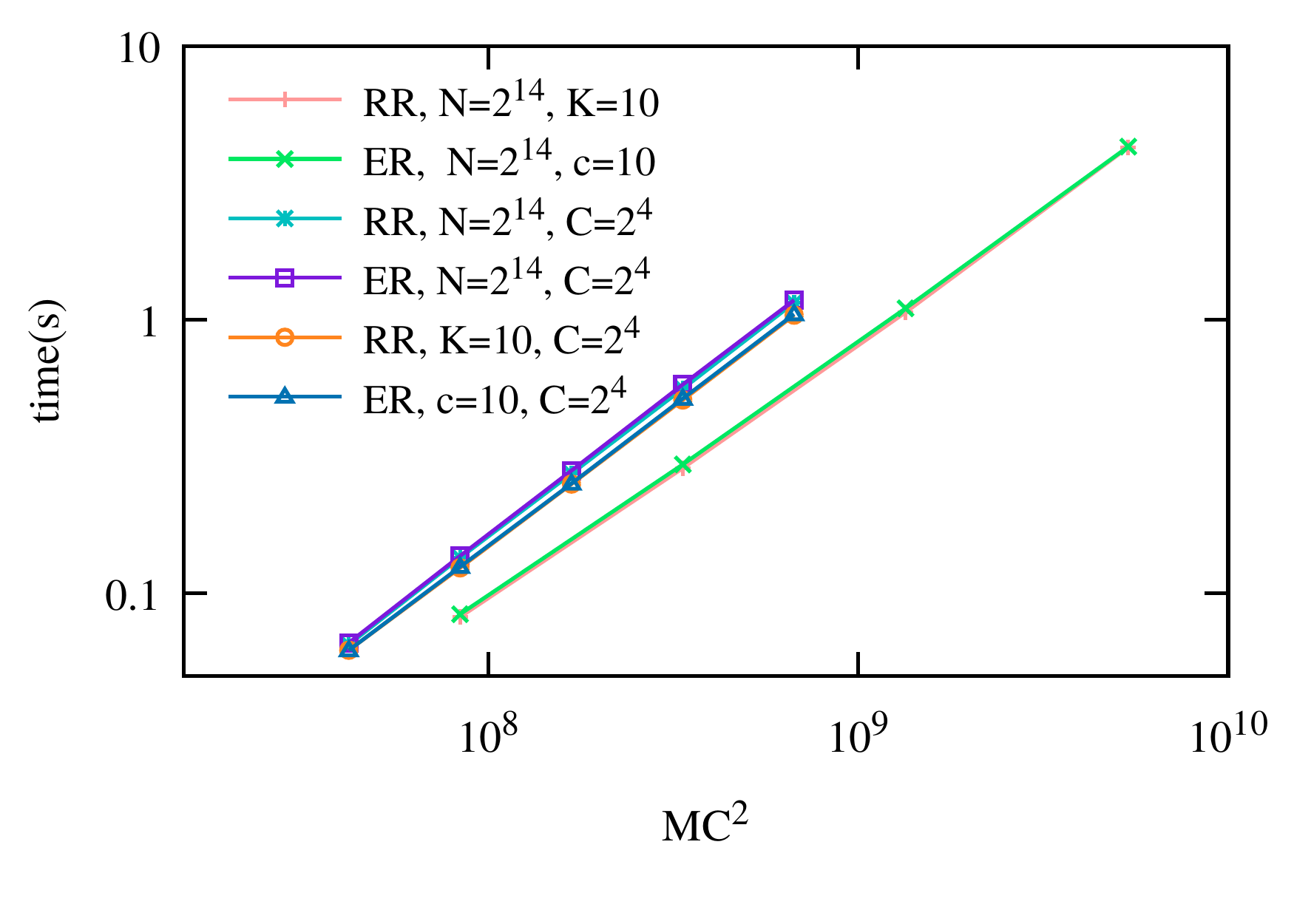}
  \caption{(Color online) The real computation time of updating all messages on a random graph. All messages are updated synchronously and parallelly in a desktop computer (AMD-2700X, dual channel memory at 2666MHz). When $N=2^{14}$ and $K=10$ in RR graph (pluses) or $c=10$ in ER graph (crosses), we run $C=2^5, 2^6, 2^7, 2^8$. When $N=2^{14}$ and $C=2^4$, we run $K=20, 40, 80, 160, 320$ in RR graph (stars) and $c=20, 40, 80, 160, 320$ in ER graph (squares). When $K=10$ in RR graph (circles) or $c=10$ in ER graph (triangles) and $C=2^4$, we run $N=2^{15}, 2^{16}, 2^{17}, 2^{18}, 2^{19}$.
    \label{fig:time}
  }
\end{figure}

The complexity of the operation $\otimes$ is in the order of $(C+1)^2$.
For a vertex $i$ with $|\partial i|=k$, we will use operator $\otimes$  $3k-2$ times to update all messages $\{q_{i\to j}^{A_i}\}_{j\in\partial i}$ (see Appendix~\ref{sec:appA} for more details).
Therefore, the computation time of updating all messages on graph $\mathcal{G}$ will be in the order of $NKC^2$ or $MC^2$, where $K$ is the average degree of each vertex.
In Fig.~\ref{fig:time}, we present the computation time of updating all messages on various random graphs with different $C$.
The computation time is proportional to the $N$, to the mean degree $c$ in ER graph or to the degree $K$ in RR graph and to the $C^2$.
What is more, the computation time is irrelevant to the type of random graph ensembles as long as they have the same average degree.

\subsection{The dismantling problem in the large $N$ limit}
In this subsection, we will discuss the dismantling problem in the large $N$ limit, where $C\to\infty$ either and the following limitation keeps
\begin{equation}
\label{eq:limit}
\lim_{N\to\infty}\frac{C}{N}=0.01.
\end{equation}
In that case, the computation time of BP iterations will be proportional to $N^3$ and both computation time and memory usage will become unaffordable.
Therefore, we hope Eq.~\ref{eq:BP} can be simplified further.

From Fig.~\ref{fig:fig2}, we know that $\overline{q}_i$ increases exponentially with growing $A/C$ as long as $A/C>0$.
Therefore, a vertex belonging to a connected component with very large size but smaller than $C/2$ is almost impossible and can be neglected.
Under this assumption, we can introduce another discrete probability distribution with only three states: 0, $\mathbb{I}$ and $\mathbb{X}$, which mean a vertex $i$ in set $\mathcal{D}$, in a connected component with finite size, and in a connected component with infinite size larger than $C/2$.
$\hat{q}_i^0$, $\hat{q}_i^\mathbb{I}$ and $\hat{q}_i^\mathbb{X}$ are marginal probabilities of vertex $i$ in state 0, $\mathbb{I}$ and $\mathbb{X}$ respectively.
Then, in the same way, we can define messages $\hat{q}_{i\to j}^0$, $\hat{q}_{i\to j}^\mathbb{I}$ and $\hat{q}_{i\to j}^\mathbb{X}$ in the cavity graph $\mathcal{G}_{\setminus i}$.
If all neighbour vertices $j\in\partial i$ in state $\mathbb{I}$ or 0, we can add vertex $i$ to set $\mathcal{S}$ and the vertex $i$ will be in state $\mathbb{I}$.
If there is only one nearest neighbour vertex in state $\mathbb{X}$, and all other neighbour vertices are in state $\mathbb{I}$ or 0, adding vertex $i$ to set $\mathcal{S}$ will generate a new connected component with infinite size but still satisfying the limitation~\ref{eq:limit}.
However, if there are more than one neighbour vertices in state $\mathbb{X}$, the size of the new generated component will be larger than $0.01N$, which means the vertex $i$ must be in set $\mathcal{D}$.
Now, we can have the following self-consistent equations of these probability distributions:
\begin{subequations}
\label{eq:SCDBPD}
  \begin{align}
\hat{q}_{i\to j}^0&=\frac{e^{-\beta\omega_i}}{\hat{z}_{i\to j}},\\
\hat{q}_{i\to j}^\mathbb{I}&=\frac{1}{\hat{z}_{i\to j}}\prod_{k\in\partial i\setminus j}(\hat{q}_{k\to i}^0+\hat{q}_{k\to i}^\mathbb{I}),\\
\hat{q}_{i\to j}^\mathbb{X}&=\frac{1}{\hat{z}_{i\to j}}\sum_{k\in\partial i\setminus j}\hat{q}_{k\to i}^\mathbb{X}\prod_{m\in\partial i\setminus j,k}(\hat{q}_{m\to i}^0+\hat{q}_{m\to i}^\mathbb{I}),\label{eq:SCDBPDc}
\end{align}
\end{subequations}
where
\begin{eqnarray}
\hat{z}_{i\to j}&\equiv&   e^{-\beta\omega_i}+\prod_{k\in\partial i\setminus j}(\hat{q}_{k\to i}^0+\hat{q}_{k\to i}^\mathbb{I})\nonumber\\
&+&\sum_{k\in\partial i\setminus j}\frac{\hat{q}_{k\to i}^\mathbb{X}}{\hat{q}_{k\to i}^0+\hat{q}_{k\to i}^\mathbb{I}}\prod_{k\in\partial i\setminus j}(\hat{q}_{k\to i}^0+\hat{q}_{k\to i}^\mathbb{I}).
\end{eqnarray}

And the marginal probability of each vertex $\hat{q}_i$ can also be computed by
\begin{subequations}
\label{eq:SCDRS}
  \begin{align}
\hat{q}_i^0&=\frac{e^{-\beta\omega_i}}{\hat{z}_i},\\
\hat{q}_i^\mathbb{I}&=\frac{1}{\hat{z}_i}\prod_{j\in\partial i}(\hat{q}_{j\to i}^0+\hat{q}_{j\to i}^\mathbb{I}),\\
\hat{q}_i^\mathbb{X}&=\frac{1}{\hat{z}_{i\to j}}\sum_{j\in\partial i}\hat{q}_{j\to i}^\mathbb{X}\prod_{k\in\partial i\setminus j}(\hat{q}_{k\to i}^0+\hat{q}_{k\to i}^\mathbb{I}),\label{eq:SCDRSc}
\end{align}
\end{subequations}
where
\begin{equation}
\hat{z}_i\equiv e^{-\beta\omega_i}+\bigg(1+\sum_{j\in\partial i}\frac{\hat{q}_{j\to i}^\mathbb{X}}{\hat{q}_{j\to i}^0+\hat{q}_{j\to i}^\mathbb{I}}\bigg)\prod_{j\in\partial i}(\hat{q}_{j\to i}^0+\hat{q}_{j\to i}^\mathbb{I}).
\end{equation}

Comparing the equations above with the BP iterations in the FVS spin-glass model discussed in \cite{PhysRevE.94.022146,Mugisha2016,Zhou2013}, we find there is only one very small difference between them: Eq.~\ref{eq:SCDBPDc} and \ref{eq:SCDRSc} use $\hat{q}_{j\to i}^\mathbb{X}$ instead of $\hat{q}_{j\to i}^\mathbb{X}+\hat{q}_{j\to i}^\mathbb{I}$.
Actually, both $\hat{q}_{i\to j}^\mathbb{I}$ in the Eq.~\ref{eq:SCDBPD} and $q_{i\to j}^i$ in the Eq. 20 of \cite{Zhou2013} are very small in their respective iteration equations.
This result confirms the connection between the dismantling problem and the FVS problem under the thermodynamic limit and explains why the lower bounds of the FVS problem and $\rho_\infty$ are close to each other and why decycling algorithms work so well in dismantling problems.

\begin{figure*}
  \includegraphics[width=1\textwidth]{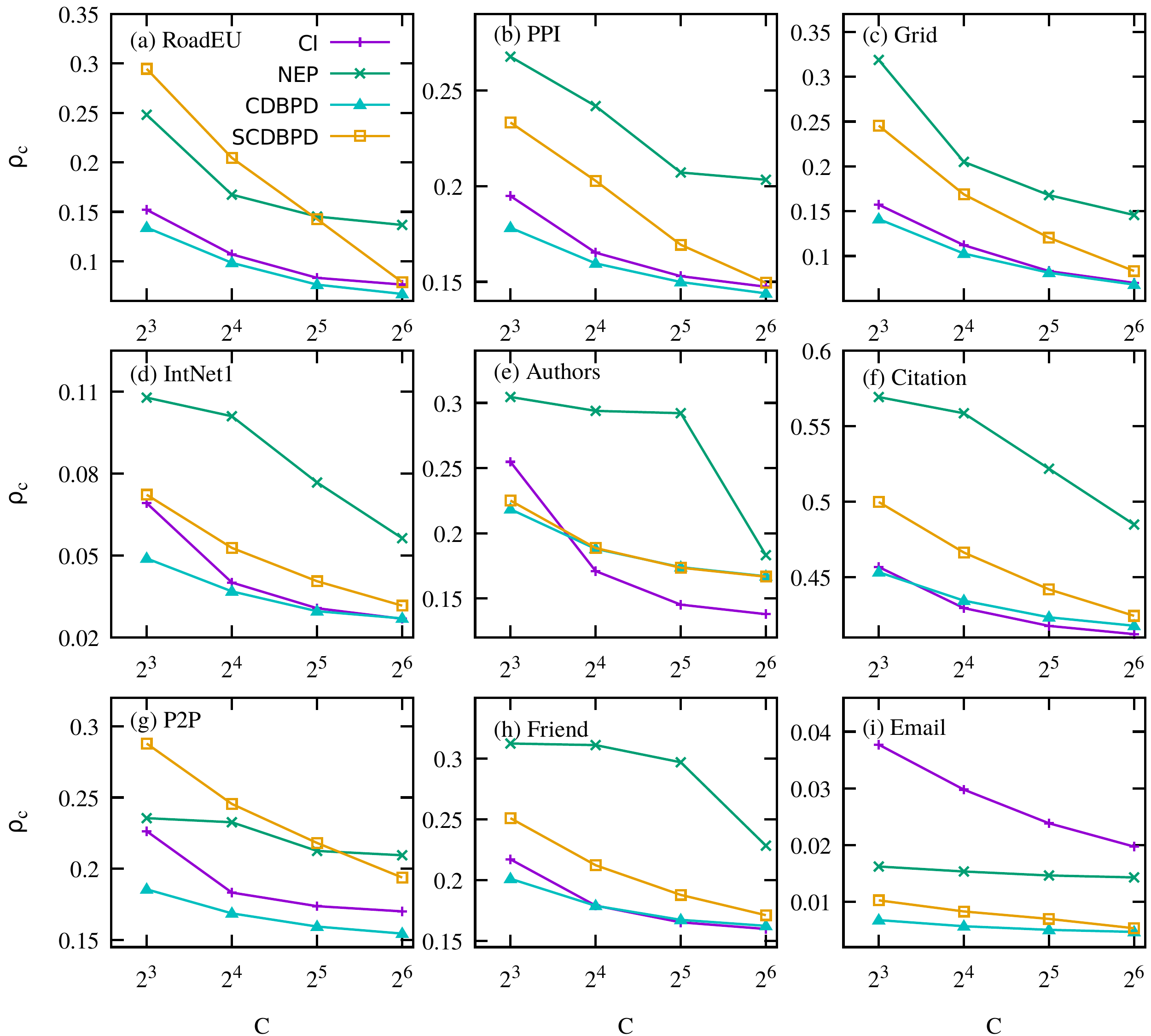}
  \caption{ (Color online) The relative size $\rho$ of set $\mathcal{D}$ for some real world networks as a function of $C$ given by CI (pluses), NEP (crosses), CD-BPD (triangles) and SCD-BPD (squares) algorithms. The numbers in the brackets are the  clustering coefficients of corresponding graph.  (a) RoadEU (0.0671) \cite{Subelj2011}, (b) PPI (0.1301) \cite{Bu2003}, (c) Grid (0.0801) \cite{Watts1998}, (d) IntNet1 (0.2522) \cite{Leskovec2005Graphs}, (e) Authors (0.6334) \cite{Leskovec:2007:GED:1217299.1217301}, (f) Citation (0.2848) \cite{Leskovec2005Graphs}, (g) P2P (0.0055) \cite{Leskovec2005Graphs}, (h) Friend(0.2367) \cite{Leskovec2005Graphs}, (i) Email (0.0671) \cite{Leskovec:2007:GED:1217299.1217301}.
  \label{fig:realNet}
  }
\end{figure*}
\subsection{The CD-BPD algorithm and SCD-BPD algorithm}
In this subsection, we develop two belief-propagation-guide decimation (BPD) algorithms based on the Eq. \ref{eq:BP} (denoted as CD-BPD) and \ref{eq:SCDBPD} (denoted as SCD-BPD) respectively to solve the CD problem on a certain graph.
The detail and pseudocode of two BPD algorithms are explained in Appendix~\ref{sec::appB}.
We compare the performance of two BPD algorithms with two other heuristic methods (collective influence algorithm with ball radius $\ell =2$ (CI)~\cite{Morone2015} and node explosive percolation algorithm with the second score definition in~\cite{Clusella2016} (NEP)) on ER, RR and SF graphs with various average degree and present all results in Fig.~\ref{fig:fig1} and~\ref{fig:fig4}.
In Fig.~\ref{fig:fig1}, we find CD-BPD gives near optimal $\mathcal{D}$ very close to the result predicted by the RS mean-field method and it is far better than other algorithms.
Because the approximation of Eq.~\ref{eq:SCDBPD} holds only in the large $C$ limit, it is not surprise that the results of SCD-BPD approach to that of CD-BPD gradually with growing $C$ and even outperform CD-BPD a little in the RR graph with $C=512$.
Actually, if $m,n\in\partial i$  and they are in the same loop with the length shorter than $C$, it is possible that they are also in the same connected component and $A_m=A_n$.
In that case, the Bethe-Peierls approximation is invalid and extra vertices will be added to $\mathcal{D}$.
The random graphs used in our computation have finite number of vertices, so the length of typical cycles is also limited (in the order of $\ln N$) \cite{Marinari2004,0295-5075-73-1-008,1742-5468-2005-06-P06005}.
Therefore, in the case of $C$ larger than the length of typical cycles, the SCD-BPD may surpass CD-BPD not only in computation efficiency but also in computation results.

Figure~\ref{fig:fig4} concerns the performance of these algorithms in all kinds of artificial random graphs.
We can see that the CD-BPD still gives the best results for the CD problem with $C=64$.
Actually, as $\rho_c$ approaches to $\rho_\infty$ in the order of $C^{-1}$, the results given by CD-BPD with $C=64$ is already very close to the $\rho_\infty$ and also close to the result given by SCD-BPD in the dismantling problem.

At last, we apply these algorithms in some real world networks, which contains plenty of communities, local loops and hierarchical levels.
The value of $\rho$ with various $C$ are presented in Fig.~\ref{fig:realNet}.
Except for the Authors, Citations and Friends networks, where CI obtains better results when $C>8$, the CD-BPD gives the minimum $\mathcal{D}$ in all tested algorithms.
We find the clustering coefficients of Authors, Citations and Friends networks are relatively large, which means there exist a mass of short loops in them.
We believe that is the main reason why CD-BPD does not work well in these instances.
Another network with conspicuous clustering coefficient is the IntNet1, where CI performs as well as the CD-BPD when $C\geq 32$.
The results of SCD-BPD also approach to that of CD-BPD with growing $C$ in these networks.

\section{Conclusion and Discussion}
In this paper, we propose a spin-glass model for the CD problem and study its properties by the RS mean-field method.
We also develop two BPD algorithms to solve the CD problem in a certain network.
The CD-BPD gives the best result in all state-of-art algorithms for the CD problem with small $C$ and SCD-BPD consumes less computation resource when $C$ is large.
Both of them work well for the large $C$ as long as the clustering coefficients of the network is small.

Although the loops in random graph may lead to negative effect to the CD spin-glass mode, the value of $\rho_c$ and $\rho_\infty$ are not overestimated because the number of loops with finite length in a random graph is also finite even in the thermodynamic limit \cite{Marinari2004,0295-5075-73-1-008,1742-5468-2005-06-P06005}.
On the other hand, the length of typical loops in random graph will be larger than $C$ as $N\to\infty$ and the RS mean-field method neglects these possible long-range correlations.
Considering both aspects together, we say CD spin-glass model provides lower bounds $\rho_c$ and $\rho_\infty$ of the CD and dismantling problem in the random graphs.

Additionally, our CD spin-glass model connects the vertex cover problem with $C=1$ and the FVS problem with $C\to\infty$.
We also notice that FVS problem shares many common features with the vertex cover problem, like the critical temperatures change with the mean vertex degree in a nonmonotonic way \cite{PhysRevE.94.022146, Zhang2009}.
So we speculate that the first-step replica-symmetry-breaking phase transition of the CD problem will belong to the same universal class with the FVS and vertex cover problem \cite{Mezard2001,MezardM2009,Monasson1995Structural}.
We will confirm this speculation in a separate paper.

The SCD-BPD algorithm solves the CD and dismantling problems in a more straightforward way than the FVS-BPD algorithm discussed in \cite{Mugisha2016}, although they have similar message passing equations.
There are three stages in decycling algorithms \cite{Braunstein2016, Mugisha2016}: finding the minimum FVS, breaking the remaining tree and introducing some possible cycles.
The SCD-BPD algorithm can give the dismantling set directly without the latter two stages.

Our numerical computations on real world networks reveal that the CD-BPD and SCD-BPD do not work well in networks with plenty of short loops.
This drawback can be made up by considering the effects of local short loops.
We can start this work from the simplest triangle structure and then extend to more complex local structures.
We believe that better results for these real world networks can be obtained in our future works.

The computations in this paper are carried out in the HPC Cluster of ITP-CAS.
We would like to thank Hai-Jun Zhou for helpful discussions and constructive suggestions.
This paper is supported by the 2016 Fundamental Research Funds for the Central Universities, Civil Aviation University of China (Grant no. 3122016L010), and the National Natural Science Foundation of China (Grant nos. 11705279 and 61503385).

\appendix

\section{Pseudocode of update all messages on graph $\mathcal{G}$ by equations \ref{eq:otimes} and \ref{eq:newBP} }
\label{sec:appA}
Here we explain how to update all messages on graph $\mathcal{G}$ by equations~\ref{eq:otimes} and \ref{eq:newBP}.
For each vertex, if we update all messages $\{q_{i \to j}^{A_i}\}_{j\in\partial i}$ together, we only need to use the $\otimes$ operation $3k-2$ times, where $k=|\partial i|$. Then the computation can be simplified further.

\begin{algorithm}[H]
  \algsetup{indent=1em}
  \caption{Update all messages on graph $\mathcal{G}$ by equations \ref{eq:otimes} and \ref{eq:newBP}.}
  \begin{algorithmic}
  \STATE Generate a random order of all vertices $\mathcal{V}$: $\{t_1,t_2,\cdots,t_N\}$;
    \FOR{r=1, $\cdots$, N}
    \STATE Select the vertex $i$ in the order of $\{t_1,t_2,\cdots,t_N\}$, then $i=t_r$;
    \STATE Define a sequence $\{j_1,j_2,\cdots,j_k\}$ for the nearest neighbours $\partial i$;
    \STATE Set message $q'_{j_1}=\{1,0,\cdots,0\}$ and $q''_{j_k}=\{1,0,\cdots,0\}$;
        \FOR{s=2, $\cdots$, k}
            \STATE Compute $q'_{j_s}=q'_{j_{s-1}}\otimes q_{j_{s-1}\to i}$;
        \ENDFOR
        \FOR{s=k-1, $\cdots$, 1}
            \STATE Compute $q''_{j_s}=q''_{j_{s+1}}\otimes q_{j_{s+1}\to i}$;
        \ENDFOR
        \FOR{s=1, $\cdots$, k}
            \STATE Compute $\tilde{q}_{i\to j_s}=q'_{j_s}\otimes q''_{j_s}$;
            \STATE Update $q_{i\to j_s}$ from $\tilde{q}_{i\to j_s}$ by Eq.~\ref{eq:newBP};
        \ENDFOR
    \ENDFOR
    \label{code:codeBP}
  \end{algorithmic}
\end{algorithm}
\section{CD-BPD algorithm and SCD-BPD algorithm}
\label{sec::appB}

For a given graph $\mathcal{G}$, BP equations cannot only estimate the size of the minimum set $\mathcal{D}$, but also give a near optimal solution of the CD problem by the BPD algorithm.
In this paper, we introduce BP equations to study the properties of CD problem with finite $C$ and then simplify it in the large $C$ limit.
Therefore, we develop two different BPD algorithms based on the original BP equations and the simplified BP equations, which are denoted as CD-BPD and SCD-BPD respectively.

At the beginning of each BPD algorithm, we set $\beta>\beta^*$ and randomly initial  all messages $\{q_{i\to j}\}$ or $\{\hat{q}_{i\to j}\}$ on graph $\mathcal{G}$. Then we empty the set $\mathcal{S}=\emptyset$ and $\mathcal{D}=\emptyset$.
In each round of the BPD algorithms, the BP equations \ref{eq:BP} or \ref{eq:SCDBPD} are performed enough times so that every message can spread its information to the entire connected component.
The BP equations may not reach its fixed point in the BPD algorithm, but it does not prevent us from computing marginal probability of each vertex $q_i^0$ or $\hat{q}_i^0$ by Eq. \ref{eq:RS} or \ref{eq:SCDRS}.
Then a small fraction of vertices with the largest $q_i^0$, $\hat{q}_i^0$ are added into set $\mathcal{D}$.
At the same time, we will also remove these vertices and their adjacent edges from the graph.
During this process, it is possible that the remaining graph breaks to many connected components, some of which will be equal to or smaller than $C$.
Because these small connected components satisfy the constraint of the CD problem, we can remove the entire connected component away from the remaining graph to avoid unnecessary vertex attacking.
The vertices in these small connected components are added to the set $\mathcal{S}$.
After that, we can iterate BP equations for the next round until all vertices are removed from the graph and $\mathcal{D}\bigcup\mathcal{S}=\mathcal{V}$. Here we present the pseudocode of two BPD algorithms in the following:

\begin{algorithm}[H]
  \algsetup{indent=1em}
  \caption{The CD-BPD and SCD-BPD algorithm based on the original BP equations and the simplified BP equations.}
  \label{alg:BPD}
  \begin{algorithmic}
    \STATE For a graph $\mathcal{G}(\mathcal{V},\mathcal{E})$ with $N=|\mathcal{V}|$ vertices, inverse temperature $\beta$, a small fraction $f$, iteration number $T$, initial all messages $\{q_{i\to j}\}$ or $\{\hat{q}_{i\to j}\}$ on the graph $\mathcal{G}$ randomly;
    \STATE Empty the set $\mathcal{D}$ and $\mathcal{S}$: $\mathcal{D}=\emptyset$, $\mathcal{S}=\emptyset$;
    \WHILE {$\mathcal{D}\bigcup\mathcal{S}\neq\mathcal{V}$}
    \FOR{t=1, $\cdots$, T}
            \STATE Try to find out the solution of BP equations by updating messages $\{q_{i\to j}\}$ or $\{\hat{q}_{i\to j}\}$ ;
    \ENDFOR
    \FOR{i=1, $\cdots$, N}
        \STATE Compute the value of $q_i^0$ or $\hat{q}_i^0$ by their corresponding RS cavity equations;
    \ENDFOR
    \FOR{s=1, $\cdots$, $f$N}
        \STATE Find the vertex $i$ with the largest $q_i^0$ or $\hat{q}_i^0$ in the remaining graph;
        \STATE Add vertex $i$ into set $\mathcal{D}$;
        \STATE Delete vertex $i$ with its adjacent edges from the remaining graph;
        \IF {There are connected components with the size not larger than $C$ in the remaining graph}
            \STATE Add all vertices in these connected components into set $\mathcal{S}$;
            \STATE Remove the entire connected components from the remaining graph;
        \ENDIF
    \ENDFOR
    \ENDWHILE
  \end{algorithmic}
\end{algorithm}
\bibliography{dismantleGroup}
\end{document}